%% file: paper.tex
\documentclass[sigplan,screen]{acmart}
\usepackage{enumitem} 
\usepackage{balance}

\author{Baijun Cheng}
\affiliation{%
  \institution{Peking University}
  \city{}
  \country{China}
}
\email{prophecheng@stu.pku.edu.cn}

\author{Kailong Wang}
\authornote{Corresponding authors}
\affiliation{%
 \institution{Huazhong University of Science and Technology}
 \city{}
 \country{China}
}
\email{wangkl@hust.edu.cn}

\author{Ling Shi}
\affiliation{%
  \institution{Nanyang Technological University}
  \city{}
  \country{Singapore}
}
\email{ling.shi@ntu.edu.sg}

\author{Haoyu Wang}
\affiliation{%
 \institution{Huazhong University of Science and Technology}
 \city{}
  \country{China}
}
\email{haoyuwang@hust.edu.cn}

\author{Yao Guo}
\authornotemark[1]
\affiliation{%
  \institution{Peking University}
  \city{}
  \country{China}
}
\email{yaoguo@pku.edu.cn}

\author{Ding Li}
\affiliation{%
  \institution{Peking University}
  \city{}
  \country{China}
}
\email{ding_li@pku.edu.cn}

\author{Xiangqun Chen}
\affiliation{%
  \institution{Peking University}
  \city{}
  \country{China}
}
\email{cherry@sei.pku.edu.cn}

\begin{abstract}

Pointer analysis has been studied for over four decades.
However, existing frameworks continue to suffer from the propagation of incorrect facts. 
A major limitation stems from their insufficient semantic understanding of code, resulting in overly conservative treatment of user-defined functions. 
Recent advances in large language models (LLMs) present new opportunities to bridge this gap.
In this paper, we propose LMPA (LLM-enhanced Pointer Analysis), a vision that integrates LLMs into pointer analysis to enhance both precision and scalability. 
LMPA identifies user-defined functions that resemble system APIs and models them accordingly, thereby mitigating erroneous cross-calling-context propagation. 
Furthermore, it enhances summary-based analysis by inferring initial points-to sets and introducing a novel summary strategy augmented with natural language.
Finally, we discuss the key challenges involved in realizing this vision.

\end{abstract}



\setcopyright{acmlicensed}
\acmDOI{10.1145/3759425.3763393}
\acmYear{2025}
\copyrightyear{2025}
\acmISBN{979-8-4007-2148-9/25/10}
\acmConference[LMPL '25]{Proceedings of the 1st ACM SIGPLAN International Workshop on Language Models and Programming Languages}{October 12--18, 2025}{Singapore, Singapore}
\acmBooktitle{Proceedings of the 1st ACM SIGPLAN International Workshop on Language Models and Programming Languages (LMPL '25), October 12--18, 2025, Singapore, Singapore}
\received[accepted]{2025-08-08}

\begin{document}

\title{Enhancing Semantic Understanding in Pointer Analysis using Large Language Models}

\maketitle




\input{sections/1.Introduction}

\input{sections/2.Background}
\input{sections/3.Approach}

\input{sections/4.Summary}

\bibliographystyle{ACM-Reference-Format} 
\balance
\bibliography{sections/reference}

\end{document}

%% file: sections/1.Introduction.tex
\section{Introduction}



In the software development lifecycle, a series of program analysis tasks (e.g., value flow analysis~\cite{PSVFA, SVF}) is typically employed to detect potential bugs and ensure software reliability. 
Among these tasks, pointer analysis serves as a foundational technique that provides critical information for other program analyses, including but not limited to variable pointing relations and aliasing relationships between variables. 
Furthermore, pointer analysis also plays a significant role in compiler optimizations~\cite{das2001estimating}.

Pointer analysis has been studied for over four decades and continues to face a fundamental challenge: striking a balance between precision and efficiency. 
In general, increasing the precision of pointer analysis through techniques such as context sensitivity and flow sensitivity often leads to higher computational costs, thereby creating a trade-off between analysis accuracy and scalability. 
To address this issue, the research community has proposed various techniques aimed at reducing the overhead of pointer analysis and making high-precision methods more practical. 
These techniques include but are not limited to selectively applying context-sensitive analysis~\cite{Zipper, Scaler, Graphick}, reducing redundant propagations~\cite{WP_DP, DEA, PUS, VSFS}, applying demand-driven approaches~\cite{SUPA, SUPA_extend, DDPA, boomerang}, and utilizing summary-based methods~\cite{pinpoint, Falcon, Saturn, Calysto}.

Although these state-of-the-art approaches reduce the analysis overhead by minimizing the unnecessary propagation of pointer information, they still exhibit limitations, particularly in the propagation of incorrect facts.
Specifically, we observe that:

\begin{enumerate}[leftmargin=*]
\item \textbf{L1: Limitations in User-Defined API Abstraction}. Many user-defined functions replicate the functionality of standard system APIs. If such functions can be abstracted in the same manner as system APIs, it becomes possible to avoid propagating pointer information into their bodies. 
Instead, the propagation can be confined to their respective caller functions, thereby preventing incorrect cross-calling-context propagation and further improving efficiency.

\item \textbf{L2: Limitations in Summarizing Complex Input Structures}. 
In real-world programs, input parameters are often typed by complex data structures such as structs or arrays, where a field of an object may point to another object. 
Existing summary approaches~\cite{pinpoint, Falcon} commonly assume that each input parameter corresponds to a single virtual object, overlooking potential internal pointer relationships among fields. 
This abstraction can lead to the loss of important aliasing and points-to information.

\item \textbf{L3: Inefficient Summary Strategy}.
Existing summary techniques typically fall into two categories: (1) recording store/load operations or MOD/REF summaries that are later inlined into callers~\cite{Calysto, Saturn}, which entails a trade-off between analysis precision and efficiency — higher precision generally requires more extensive cloning; or (2) applying code transformations~\cite{pinpoint, Falcon} to expose side effects at the source level, thereby obviating the need for explicit summaries. However, the latter approach tends to be less effective in the presence of intricate or irregular side-effect patterns.

\end{enumerate}

The limitations mentioned above are inherently semantic and are difficult to address using traditional heuristic-based techniques. 
Recent advances in large language models (LLMs) offer a promising new perspective for tackling these issues. 
LLMs have demonstrated impressive capabilities across a range of program-language and natural-language tasks, including code summarization~\cite{CodeSummarization}, code translation\\\cite{CodeTranslation}, and code generation~\cite{CodeGeneration} by capturing rich semantic information from code. 
In the domain of taint analysis, recent studies~\cite{RepoAudit, InferROI, GPTScan, Lara, IRIS, Artemis} further indicate that LLMs are capable of reasoning about deep program semantics, such as identifying whether a function serves as a source or sink, and even generating potential taint propagation paths. 

To mitigate the limitations above, we propose a vision for a Large Language Model-enhanced Pointer Analysis framework (LMPA). 
LMPA leverages the semantic understanding capabilities of large language models to enhance traditional pointer analysis along several dimensions:

\begin{itemize}[leftmargin=*]
\item To address limitation L1, LMPA utilizes LLMs to identify user-defined functions that semantically resemble standard system functions such as memcpy, snprintf, malloc, or free. 
These functions can then be abstracted and modeled accordingly to avoid unnecessary propagation into their bodies.

\item To address limitation L2, LMPA improves summary construction by analyzing both the type structure of input parameters and the function’s code. 
For parameters that are arrays or pointers to structs, LMPA examines their internal fields and leverages LLM-based code analysis to infer which fields are likely to be \text{NULL} or point to other virtual objects. 
This enables the construction of a more precise virtual initial points-to set for parameters.

\item To address Limitation L3, LMPA introduces a natural language enhanced summary strategy that differs from traditional bottom-up pointer analysis methods. It uses LLMs to generate a natural language description of a function’s side effects, which serves as its summary. 
When the analysis reaches the caller, this summary is decoded by the LLM and merged into the caller’s analysis state. 
Compared to conventional summaries, natural language offers a more customized and flexible representation, capturing path- and flow-sensitive behaviors while reducing the complexity–precision trade-off common in traditional approaches.
\end{itemize}

This paper advocates for collaboration between the academic and industrial communities to address the challenges involved in implementing LMPA. 
The ultimate goal is to leverage the code understanding capabilities of LLMs to overcome limitations that are difficult to address within traditional pointer analysis frameworks, thereby improving both the precision and scalability of existing analysis techniques.

%% file: sections/2.Background.tex
\section{Background and Related Work}

\subsection{LLM for static analysis}

LLMs have been increasingly applied to static analysis, such as call graph construction~\cite{SEA}, with most efforts focusing on vulnerability detection. 
Several studies leverage LLMs to enhance traditional tools, while others apply LLMs directly for vulnerability identification.

In the first category, LLift~\cite{LLift} leverages LLMs to analyze complex constraints along vulnerability paths, offering a lightweight alternative to symbolic execution and significantly reducing false positives in tools like UBITech~\cite{UBITech}. 
LATTE~\cite{Latte} employs LLMs to automatically identify taint sources and sinks introduced by external libraries in binary programs, addressing traditional static analysis limitations when handling third-party code. 
IRIS~\cite{IRIS} uses LLMs to recognize taint sources and sinks from third-party libraries and assists in interpreting taint propagation reports generated by CodeQL~\cite{codeql}, reducing both false positives and false negatives.
InferROI~\cite{InferROI} identifies user-defined resource allocation and deallocation functions to simplify resource-leak modeling, transforming complex interprocedural detection into more efficient intra-procedural analysis. GPTScan~\cite{GPTScan} detects potentially vulnerable operations in smart contracts using LLMs and enhances reliability through static reachability analysis. 
Lara~\cite{Lara} improves taint analysis for IoT firmware by identifying external-input functions, while Artemis~\cite{Artemis} uses LLMs to understand third-party PHP libraries and detect injection vulnerabilities by identifying user input sources and injection points.

In the second category, Du et al.~\cite{Vul-RAG} observed that code with similar vulnerability patterns often recurs in real-world software. Based on this, they proposed Vul-RAG, a retrieval-augmented generation (RAG) method that stores summarized vulnerable/fixed code from the training set in a database. At detection time, it retrieves similar cases by summarizing and querying the target code.
Similarly, Yang et al.~\cite{Knighter} proposed KNighter, which differs by generating and storing checkers from training examples instead of summaries. These checkers are then directly applied to detect similar bugs in the target code.

\subsection{Pointer Analysis}

Recent research on pointer analysis has primarily focused on reducing unnecessary computational overhead to improve the scalability of more precise analysis techniques. Common strategies include selective context-sensitive analysis, demand-driven methods, redundancy elimination, and summary-based approaches.

Graphick~\cite{Graphick} introduces a learning-based selective context-sensitivity strategy that outperforms existing approaches such as Zipper~\cite{Zipper} and Scaler~\cite{Scaler} in identifying critical program regions for context-sensitive analysis. 
PUS~\cite{PUS} reduces the average analysis scope of each iteration in Andersen’s algorithm~\cite{Andersen} to just 4\% of its original size by computing a causal graph that captures essential propagation dependencies. 
VSFS~\cite{VSFS} significantly lowers the cost of flow-sensitive pointer analysis by mapping different value-flow graph nodes to the same version, thereby reducing redundant computations.
In addition, Falcon~\cite{Falcon} implements a summary-based, path-sensitive pointer and data dependence analysis framework that achieves favorable trade-offs between precision and performance.

\begin{figure}[t]
  \centering
  \includegraphics[width=0.45\textwidth]{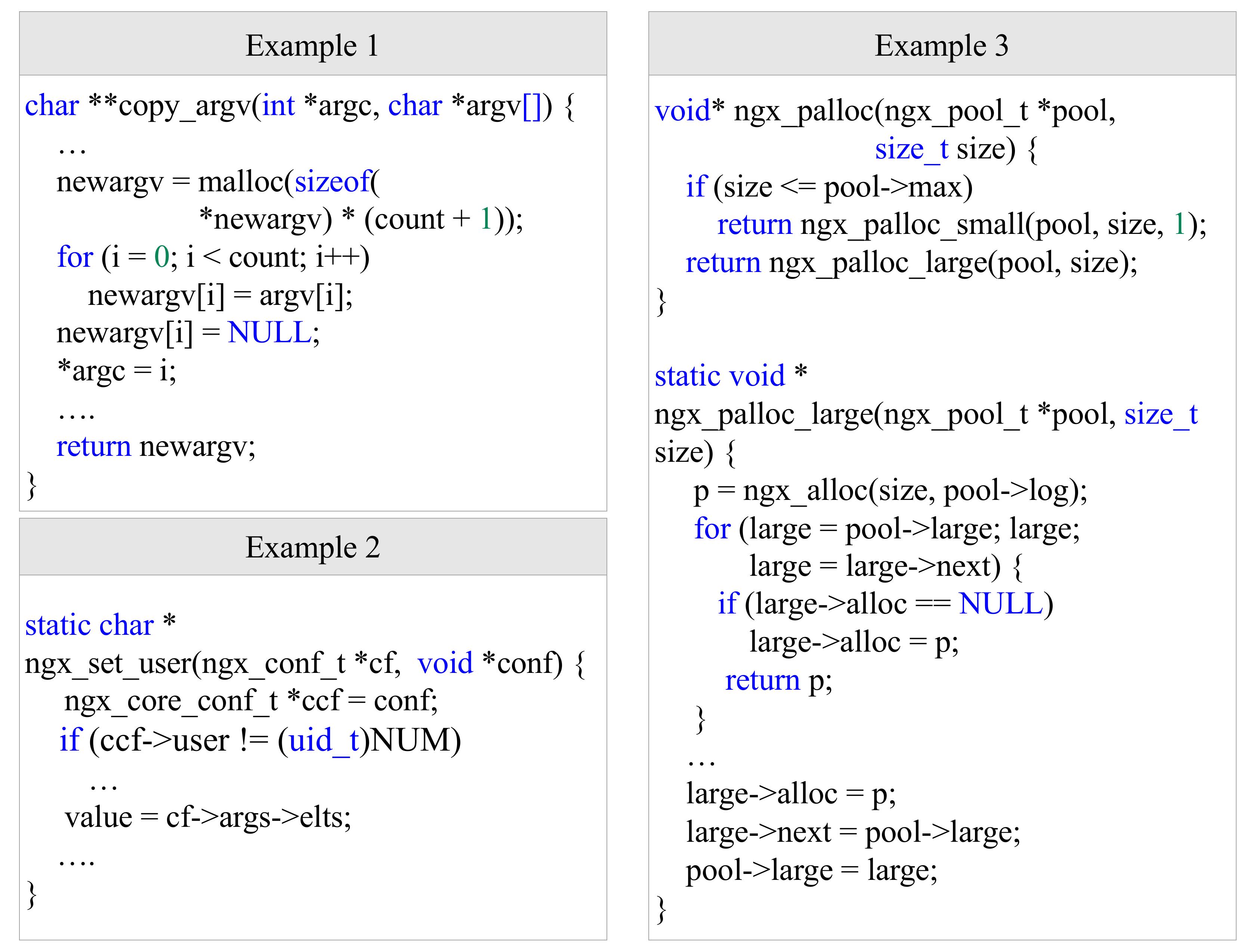}
  \caption{Examples of existing analysis limitations.}
  \label{fig:example}
\end{figure}

While these state-of-the-art techniques have significantly reduced the overhead of pointer analysis and improved the feasibility of adopting more precise analyses, the propagation of incorrect facts remains a persistent issue. 
We observe that a substantial portion of such inaccuracies is related to cross-calling context propagation.
We illustrate this problem in Figure~\ref{fig:example}, where:

\begin{enumerate}[leftmargin=*]
\item In Example 1, the user-defined function \texttt{copy\_args} behaves similarly to a combination of system API \texttt{malloc} and \texttt{memcpy}.
Existing analysis frameworks treat their callers conservatively, causing the return value \texttt{newargv} to imprecisely point to the argument \texttt{argv} of all callers. 
As a result, each caller may incorrectly receive \texttt{argv} values from others.

\item In Example 2, both parameters of \texttt{ngx\_set\_user} are struct pointers, and their internal fields are accessed within the function body. 
However, existing summary approaches typically abstract each parameter point to a distinct virtual object, without considering whether their fields may point to other objects. 
As a result, the points-to sets of fields such as \texttt{value} and \texttt{ccf->user} are imprecisely treated as \texttt{NULL}.

\item  In Example 3, \texttt{ngx\_palloc} exhibits input-dependent behaviors, both involving side effects. 
Traditional summary approaches incur substantial overhead even when encoding path-insensitive information via cloning, and tend to propagate imprecise or incorrect facts when inlining such summaries into callers. 
Moreover, the side effects of \texttt{ngx\_palloc\_large}, which involve complex linked list manipulations, are difficult to expose to callers through code transformation techniques.

\end{enumerate}

The aforementioned limitations are fundamentally rooted in the lack of code semantic understanding, making them difficult or prohibitively expensive to address using traditional analysis techniques. 
Recent advances in LLMs offer new opportunities to tackle these challenges through learned code reasoning. 
Motivated by this, we propose a vision for an LLM-enhanced Pointer Analysis framework~(LMPA).

%% file: sections/3.Approach.tex
\section{LMPA: Methodology And Challenges}

\begin{figure}[t]
  \centering
  \includegraphics[width=0.45\textwidth]{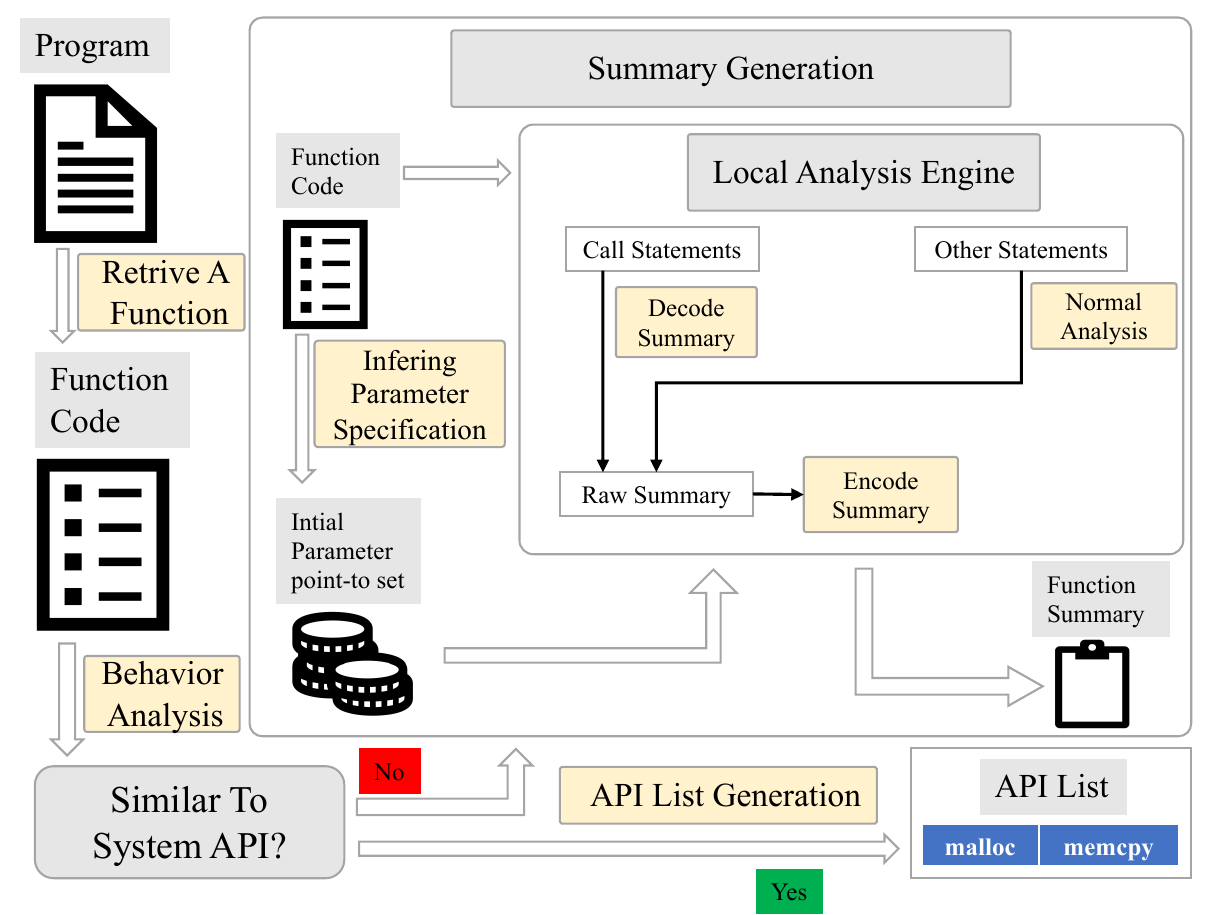}
  \caption{Overview of LMPA.}
  \label{fig:approach}
\end{figure}

Figure~\ref{fig:approach} illustrates the methodology of LMPA. 
Following prior work~\cite{pinpoint, Falcon, Saturn, Calysto}, we adopt a bottom-up, summary-based analysis framework, where functions are analyzed in reverse topological order along the call graph to incrementally build pointer analysis summaries.
For each function, (1) LMPA first performs \textbf{behavior analysis} to determine whether its semantics resemble known system APIs. 
If so, it generates an equivalent system API list, which serves as a lightweight summary for that function. 
If not, LMPA proceeds to construct a more detailed summary using an LLM-enhanced approach~(\textbf{summary generation}).
Specifically, (2) LMPA begins by analyzing the types, struct relationships, and in-function usage patterns of input parameters to infer initial parameter specification, and initializes their points-to sets accordingly.
(3) During intra-procedural analysis, standard Andersen-style rules are applied to most statements. 
For call statements, after retrieving the callee’s natural language-based summary, LMPA first invokes the LLM to decode it into a format suitable for pointer analysis. As the summary may already encode partial path- and flow-sensitive information, standard path- and flow-sensitive rules are not required. 
After completing the analysis of the current function, LMPA generates a raw summary, which is then refined into a concise natural language description using the LLM. 
This final summary is intended to eliminate incorrect facts introduced by path- and flow-insensitive reasoning.
The following section elaborates on the methodology and outlines the key challenges.

\subsection{Behavior Analysis}

The goal of \textbf{behavior analysis} is to determine whether a given function can be approximately abstracted as a standard system API list. For example, in Figure~\ref{fig:example}, behavior analysis should identify that \texttt{copy\_argv} in Example 1 can be reasonably approximated by a sequence of system API calls, whereas function \texttt{ngx\_set\_user}, \texttt{ngx\_palloc\_large}, and \texttt{ngx\_palloc} cannot.

\textbf{Challenges:} \textbf{1) Limitations of heuristic-based API approximation}. System APIs encompass a wide range of categories, making it infeasible to exhaustively model all possible types. Even widely used pointer analysis tools such as SVF~\cite{SVF} focus on a limited subset, including APIs related to memory copying (e.g., \texttt{memcpy}), memory setting (e.g., \texttt{memset}), allocation, and deallocation. 
Accurately determining whether a user-defined function in real-world programs can be approximated by one or more system API types requires a certain degree of semantic code understanding. The challenge becomes even greater when a function's behavior corresponds to a combination of multiple APIs, making it difficult to characterize using heuristic rules alone.
\textbf{2) Existence of side-effects}. User-defined functions frequently exhibit side effects, such as modifying non-local memory. In particular, those designed to mimic system APIs often manipulate global state variables.
These modifications typically do not intersect with the points-to sets relevant to most pointer analysis tasks and can therefore be safely ignored. 
Nonetheless, once a user-defined function is identified as semantically similar to a set of system API calls, it becomes non-trivial to determine whether its side effects are benign (e.g., confined to global state) or may interfere with the accuracy of pointer analysis. 
This presents a key challenge in abstracting such functions without introducing unsoundness.

\subsection{API List Generation}

Once a user-defined function is determined to be abstractable as a system API list, the next step is to generate the corresponding API sequence. 
For instance, in Example 1 of Figure~\ref{fig:example}, the generated API list for \texttt{copy\_argv} would be \texttt{malloc} followed by \texttt{memcpy}. 
In most cases, this sequence can be directly extracted from the results of the preceding behavior analysis.

\subsection{Inferring Parameter Specification}

The goal of this step is to generate the initial points-to sets for all parameters of a given function. 
For example, in the function \texttt{ngx\_set\_user} shown in Example 2 of Figure~\ref{fig:example}, the two parameters may initially point to virtual objects $\{o_1\}$ and $\{o_2\}$, respectively. Additionally, the example contains a field-sensitive points-to relation: $o_1.\text{args} \rightarrow  \{o_3\}$, $o_3.\text{elts} \rightarrow \{o_4\}$, $o_2.\text{user} \rightarrow \{o_5\}$.

\textbf{Challenge:} Although the pointer relationships among fields of structs or arrays can be complex, it is often unnecessary to assume that all fields point to valid objects within the scope of a single function. In many cases, these fields may simply hold \texttt{NULL} values, and eagerly initializing their points-to sets can introduce unnecessary analysis overhead. Therefore, an effective strategy is required to selectively identify fields whose points-to sets are likely to be non-empty, and to initialize only those sets accordingly.

While LLMs possess a certain degree of semantic understanding and can assist in inferring which fields require points-to set initialization, determining which portions of the code are most informative for such inference remains a non-trivial challenge and requires further analysis.

\subsection{Encode and Decode Summary}

In LMPA’s summary strategy, the encode and decode steps are central. 
For instance, in Example 3 of Figure~\ref{fig:example}, the natural language summary of \texttt{ngx\_palloc\_large} should describe that the function allocates a new heap object and inserts it into the large chunk list of the memory pool represented by the pool parameter. 
When encoding its caller \texttt{ngx\_palloc}, LMPA does not immediately decode \texttt{ngx\_palloc\_large}'s summary into concrete points-to facts. 
Instead, it retains the summary in its natural language form while augmenting it with a conditional clause: ``under the condition \texttt{size > pool->size}''. 
During analysis of callers of \texttt{ngx\_palloc}, the decoding process involves translating the natural language summary into a sequence of low-level instructions that reflect the pointer operations. 
For instance, the summary’s assertion that \texttt{ngx\_palloc\_large} returns a heap object stored in the pool must be mapped to: (1). a heap allocation effect (e.g., \texttt{p = \&o}), and (2). a store operation linking the object to the pool (e.g., \texttt{pool->large = o}).

\textbf{Challenges:} \textbf{1) Require a new form of intermediate representation}. 
In pointer analysis, the analysis results are typically represented as points-to sets, whereas the summaries in LMPA are expressed in natural language. 
While conventional static analysis encodes side-effect summaries as explicit instruction sequences, LMPA's natural language summaries present unique translation challenges. 
The key difficulty lies in establishing accurate bidirectional mapping between these instruction-level semantics and natural language descriptions - a process that requires LLMs to reconcile structural code patterns with their linguistic representations while maintaining semantic fidelity.
To address this, it may be necessary to define a novel intermediate representation that facilitates more reliable and structured translation between points-to information and natural language summaries.

\noindent\textbf{2) Atomic operations for LLM-guided summary refinement}. 
Beyond defining an intermediate representation, it is also important to introduce atomic operations over it. 
Since Andersen-style pointer analysis produces path- and flow-insensitive results, we propose using LLMs with source code context to refine these into partially sensitive summaries. 
Simple atomic operations, such as \textbf{kill}, can help eliminate spurious facts and serve as LLM-friendly primitives for summary refinement.

%% file: sections/4.Summary.tex
\section{Discussion}

While LLMs demonstrate promising potential in automated code analysis tasks, their inherent limitations introduce several challenges, including hallucinations, constrained reasoning capabilities, and output nondeterminism. 
These issues are difficult to eliminate entirely, but can be mitigated through targeted strategies.
A critical challenge arises when LLMs generate summaries for pointer analysis: hallucinations may lead to incorrectly decoded points-to sets. 
To address this, a two-pronged approach can be used. First, the LLM can self-validate its output through iterative refinement—e.g., by prompting it to reevaluate its initial results against logical constraints. 
Second, heuristic rules derived from existing pointer analysis techniques can be applied to verify the LLM’s outputs, with feedback loops to correct inconsistencies. 
While not foolproof, these methods reduce hallucination-induced errors.

Notably, when LLMs assist in points-to analysis, their outputs can be conservatively bounded by the supersets produced by traditional analyses (e.g., Andersen’s algorithm). 
However, hallucinations risk omitting critical facts, introducing unsoundness. 
For instance, in the sequence \texttt{p = malloc(); free(p);}, if the LLM fails to infer that p points to an allocated heap object at the second statement, the analysis may miss the memory leak—leading to false negatives in bug detection. 
Such inaccuracies cascade downstream: missing facts may cause false positives or false negatives, undermining the reliability of security-focused tools.

Ultimately, while hybrid approaches combining LLMs with classical analyses can alleviate these issues, fundamental limitations persist. Future work should explore tighter integration of symbolic reasoning frameworks to harden LLM outputs against unsoundness.

\section{Summary}

This paper proposes LMPA (LLM-enhanced Pointer Analysis), a framework that leverages large language models to address key limitations in traditional pointer analysis. 
LMPA improves precision and scalability by identifying user-defined functions similar to system APIs and modeling them to avoid incorrect fact propagation across calling contexts. It also enhances summary-based analysis through more accurate initial points-to sets and lightweight flow- \& path-sensitive natural language enhanced summaries.

This approach integrates traditional static analysis with large language models, opening new avenues for program analysis and vulnerability detection. We also highlight key challenges to realize this vision and outline a roadmap for future research at this intersection.

\begin{acks}

We thank the anonymous reviewers for their comments. 
This work was supported by Beijing Natural Science Foundation(L243010).

\end{acks}